\documentclass[aps,prl,reprint,superscriptaddress,nofootinbib,amsfonts,amssymb,amsmath]{revtex4-1}

\usepackage[utf8]{inputenc}

\usepackage{diagbox}

\usepackage{mathtools}
\usepackage{amsfonts}
\usepackage{mathrsfs}
\usepackage{bbm}
\usepackage{slashed}
\usepackage{titlesec}

\usepackage{graphicx}
\usepackage{color}
\usepackage{array}
\usepackage{esint}
\usepackage{placeins}
\usepackage{booktabs}
\usepackage{makecell}
\usepackage{epstopdf}
\usepackage[caption=false]{subfig}

\usepackage{xspace}
\usepackage{siunitx}
\usepackage{hyperref}
\usepackage[nameinlink]{cleveref}
\usepackage{appendix}

\usepackage{xifthen}
\usepackage{xcolor}
\hypersetup{
	colorlinks,
	linkcolor={red!75!black},
	citecolor={blue!75!black},
	urlcolor={blue!75!black}
}
\usepackage{comment}

\setkeys{Gin}{width=0.48\textwidth}

\graphicspath{
	{./figures/}
	{../figures/}
}

\def\s0#1#2{\mbox{\small{$ \frac{#1}{#2} $}}}
\def\0#1#2{\frac{#1}{#2}}

\newcommand{\fatg}{{\rm{I}}\!\!{\rm{I}}\!\!\Gamma}

\def\eq#1{\eqref{#1}}
\def\Eq#1{Eq.~\eqref{#1}}

\newcommand{\sumint}{\int\hspace{-4.8mm}\sum}
\newcommand{\imag}{\text{i}}

\usepackage{xcolor}

\begin{document}

\title{A novel quark pairing  in sQGP induced by the non-Abelian feature of the interaction
}

\author{Fei \surname{Gao}}
\email{fei.gao@bit.edu.cn}
\affiliation{School of Physics, Beijing Institute of Technology,  100081 Beijing, China}

\author{Yi Lu}

\email[Corresponding author: ]{qwertylou@pku.edu.cn}
\affiliation{Department of Physics and State Key Laboratory of Nuclear Physics and Technology, Peking University, Beijing 100871, China}

\author{Yu-xin Liu}
\email{yxliu@pku.edu.cn}
\affiliation{Department of Physics and State Key Laboratory of Nuclear Physics and Technology, Peking University, Beijing 100871, China}
\affiliation{Center for High Energy Physics, Peking University, 100871 Beijing, China}
\affiliation{Collaborative Innovation Center of Quantum Matter, Beijing 100871, China}


\begin{abstract}
We solve the coupled Dyson-Schwinger equations for quark propagator and quark gluon vertex in the Nambu-Gorkov basis which is widely applied to study the color superconductivity. After considering the non-Abelian feature in the off-diagonal part of  quark gluon vertex,  we acquire a quark pairing gap  in chiral limit   above the chiral phase transition temperature $T_c$. The gap persists  up to $2-3\,T_c$ and vanishes at higher temperature. Such a quark pairing characterizes the strongly coupled quark gluon plasma phase as a  new phase and distinct from the phase with quasi quarks and gluons.  Its new features   can be  disclosed in the heavy ion collision experiments.

\end{abstract}

\date\today

\maketitle

\emph{Introduction.--}
Due to the asymptotic freedom property of Quantum Chromodynamics (QCD),  the QCD matter at extremely high temperature is considered to be  deconfined quasi-quarks and gluons which is  characterized as the quark gluon plasma (QGP)~\cite{Gross:1980br,Rischke:2003mt,Shuryak:2003xe,Shuryak:2014zxa}.  
Naturally, a phase transition between  the hadronic phase and the QGP  phase is supposed to take place during the heavy ion collisions~\cite{Akiba:2015jwa,Luo:2017,HADES:2019auv,lovato2022long,ALICE-USA:2022glt,Arslandok:2023utm,Karsch:2022jwp}.  
The theoretical computations including the lattice QCD simulations~\cite{Aoki:2006we,Philipsen:2012nu,Ratti:2022qgf,Borsanyi:2020fev,HotQCD:2018pds}, functional QCD approaches (Dyson Schwinger equations~\cite{Roberts:2000aa,Qin:2010nq,Fischer:2014,Fischer:2018sdj,Gunkel:2021,Gao:2020fbl,Gao:2020qsj,Lu:2023mkn} and functional Renormalization group approach~\cite{Fu:2019hdw,Dupuis:2020fhh,Fu:2022gou}) and also effective models~\cite{Buballa:2003qv, Fukushima:2013rx,Schaefer:2008ax,Shao:2011fk,He:2013qq,Kojo:2020ztt,Chen:2018vty,Hippert:2023bel,Cai:2022omk} have made great progress on the chiral  phase transition of QCD,  while the deconfinement phase transition is still not very clear.

Nevertheless, it is  widely believed that the QGP matter as a new form of QCD matter, has been created in heavy ion collisions~\cite{BRAHMS:2004adc,PHENIX:2004vcz,PHOBOS:2004zne,STAR:2005gfr,Gyulassy:2004zy,Gyulassy:2003mc,Wang:1996yf}.  
The hydrodynamic simulation for the collective flow  suggests that at near  above the pseudo-critical phase transition temperature $T_c$, the QGP  behaves a nearly perfect fluid with a very small shear viscosity~\cite{Song:2008hj,Song:2010mg,Kovtun:2004de}, and therefore,  the matter  in this regime is believed to be strongly-coupled QGP (sQGP).  
Besides of the small viscosity and its fluidity property, the non-perturbative computations  predicts that  at $T\in [T_{c}, 3T_{c}]$,  
the thermodynamic quantities are  far away from the Stefan-Boltzman  free  limit. For instance, the pressure of sQGP at $T=2 T_{c}$ is $20\%$ lower than the limit~\cite{Karsch:2001vs,Borsanyi:2010cj,Bazavov:2014pvz}.  
Besides, an additional  zero mode apart from the two normal modes of wQGP is found in the quark spectral function near above $T_c$~\cite{Qin:2010pc,Gao:2014rqa,Su:2014rma, Fischer:2017kbq}.
The recent lattice QCD simulations also find that  there exists an enhanced chiral spin symmetry on this temperature domain~\cite{Denissenya:2014poa,Glozman:2022lda,Glozman:2022zpy}. The bound states in sQGP are also conjectured  to explain its peculiar properties~\cite{Shuryak:2004tx,Shuryak:2003ty}. 
All these observations indicate that sQGP is not simply a phase consists of quasi quarks and gluons. 


In this work, with a direct computation from the coupled  Dyson-Schwinger equations (DSEs) of quark propagator and quark gluon vertex in the Nambu-Gorkov basis~\cite{Buballa:2003qv,Alford:2007xm} in chiral limit,
we find that in the range of temperature near above  the $T_c$,  i.e.,  $T\in [T_{c}, 3T_{c}]$, despite of a vanishing chiral condensate as $\langle \bar{\Psi} \Psi \rangle=0$, a finite quark pairing gap of $\langle \Psi^\alpha_iC\gamma_5 \Psi^\beta_j \rangle$ emerges.  
The pairing is related to the non-Abelian feature in the vertex, which its generation mechanism is distinct from that of the color superconducting phase located at low temperature and high chemical potential~\cite{Hong:1999fh,Nickel:2006vf,Muller:2016fdr}.  
The non-Abelian feature in the quark gluon vertex gives a model independent relation between quark pairing gap $\Delta$ and the dimension-2 gluon condensate  $\langle g^{2}A^{2} \rangle$, which indicates that this quark pairing is dominant by the glue dynamics.
%
The  findings here may shed light on revealing the underlying physics of the phenomena in heavy ion collisions and deepen the understanding of QCD.

\emph{The Dyson-Schwinger equations in Nambu-Gorkov basis.--}
To study the quark pairing, we implement the Nambu-Gorkov basis which  extends the fermion field as:

\begin{equation}
\label{eq:NGbasis}
\begin{gathered}
\Psi=
\begin{pmatrix}
\psi  \\
\psi_{C}  \\
\end{pmatrix}, \qquad 
\overline{\Psi} =
\begin{pmatrix}
\overline{\psi}, \overline{\psi}_{C}   \\
\end{pmatrix},
\end{gathered}
\end{equation}
with $\psi_{C} = C\psi^{*}$ the charge-conjugator spinor obtained through the charge conjugation matrix $C = \gamma^{2} \gamma^{4}$.
The free quark propagator in the new basis is simply a doubling of the degrees of freedom and at finite temperature $T$ and quark chemical potential $\mu$, it can be written as:
\begin{equation}
\label{eq:freePropa}
\begin{gathered}
{\bold S}^{-1}_{0} (p) =
\begin{pmatrix}
\imag\gamma\cdot p - \gamma_{4} \mu,& 0  \\
0, &\imag\gamma\cdot p + \gamma_{4} \mu  \\
\end{pmatrix}
\end{gathered}
\end{equation}
with $p = (\vec{p},\omega_{m})$ the quark momentum, $\omega_m=(2m+1)\pi T$ the quark Matsubara frequency. Here the chiral limit is taken for the current quark mass. 
For convenience we also denote  $p_{\pm} =(\vec{p},p_4^\pm=\omega_m\pm i\mu)$ and $\bar{p}=(\vec{p},\bar p_{4}=2m\pi T)$ for the following discussion, with the same notations for $q$ and $k$.

The general form of the quark self energy in the Nambu-Gorkov formalism reads:
 \begin{equation}
\label{eq:SelfEnergy}
\begin{gathered}
{\bold\Sigma}(p)\equiv
\begin{pmatrix}
\Sigma_+(p),& \Phi_-(p)  \\[1mm]
\Phi_+(p),&\Sigma_-(p)  \\
\end{pmatrix},
\end{gathered}
\end{equation}
with $\mathcal{F}_-(p_4,\vec{p})=\gamma_2 \mathcal{F}_+^T(-p_4,\vec{p})\gamma_2$ for $\mathcal{F} = \Sigma,\,\Phi$. The quark propagator is then expressed as:
\begin{equation}
\label{eq:Propa}
\begin{gathered}
{\bold S}=
\begin{pmatrix}
S_+,& T_-  \\[1mm]
T_+,&S_-  \\
\end{pmatrix},
\end{gathered}
\end{equation}
with
\begin{eqnarray}
\label{eq:Propa1}
S_{\pm} \! &=& \! \left\{[S_{\pm,0}]^{-1} \! + \! \Sigma_{\pm} \! - \! \Phi_{\mp}([S_{\mp,0}]^{-1} \! + \! \Sigma_{\mp})^{-1}\Phi_{\pm} \right\}^{-1}, \quad \\
T_{\pm} \! &=& \! -([S_{\mp,0}]^{-1} +\Sigma_{\mp})^{-1}\Phi_{\pm} S_{\pm} .
\end{eqnarray}

\begin{figure}[t]
  \centering
  \includegraphics[width=0.9\columnwidth]{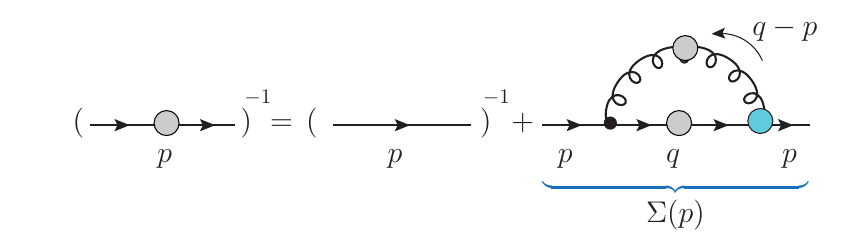}\\
  \includegraphics[width=1.1\columnwidth]{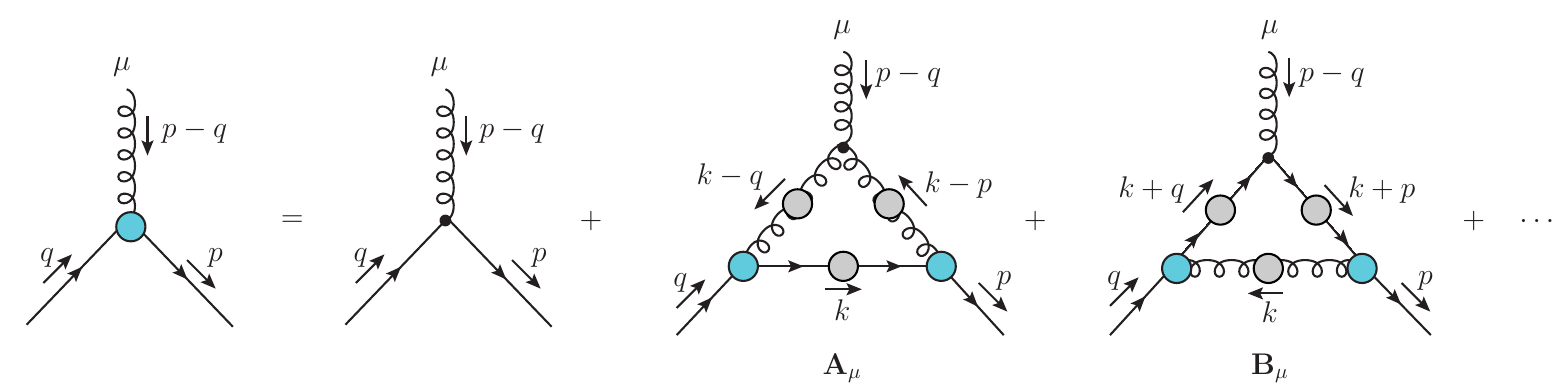}
  \caption{Schematic representation of the Dyson Schwinger equations for quark propagator and quark gluon vertex.  The one marked with $\mathrm{B}_{\mu}$ in quark gluon vertex can also be found in Abelian gauge theory, 
  while that involves the three gluon vertex (marked with $\mathrm{A}_{\mu}$ ) stands for the non-Abelian feature.} \label{fig:SDE}
\end{figure}

The quark gap equation   at finite $T$ and $\mu$ is depicted in Fig.~\ref{fig:SDE} which can be  expressed as:
\begin{eqnarray}
\label{eq:gap1}
{\bold S}(p)^{-1} &=&  {\bold S}_0^{-1}
  +{\bold \Sigma}(p) \, ,  \\[1mm]
\nonumber
{\bold\Sigma}(p) &=&\sumint_k\; {g^{2}} G^{a a^{\prime}}_{\mu\nu} ((p-k)^{2}; T, \mu)   {\bold \Gamma}^{a,0}_{\mu} \, {\bold S}(k) \,
{\bold \Gamma}^{a^\prime}_{\nu} (k, p)\, ,
\label{eq:gap2}
\end{eqnarray}
with $$\sumint_{k} = T\sum_{n=-\infty}^{\infty} \! \int\frac{d^3{\vec{k}}}{(2\pi)^{3}},$$ and the bare quark gluon vertex:
\begin{equation}\label{eq:barevtx}
{\bold \Gamma}^{a,0}_{\mu} = \frac{ \bold \Lambda^a}{2}\gamma_{\mu}, \quad
\frac{ \bold \Lambda^a}{2}= \begin{pmatrix}
\frac{\lambda^a}{2}, & 0 \\[1mm]
0, & -\frac{(\lambda^{a})^{T}}{2} \\
\end{pmatrix} ,
\end{equation}
and
$ G^{aa^{\prime}}_{\mu\nu} ((p-k)^{2}; T, \mu) $ is the dressed-gluon propagator which can be considered as color-diagonal with imaginary components~\cite{Muller:2016fdr} as:
\begin{equation} G^{a a^{\prime}}_{\mu\nu}( p^{2};T,\mu)=\delta^{a a^{\prime}}(P^{L}_{\mu\nu}( p)G_{L}(\bar p^{2})+P^{T}_{\mu\nu}(p) G_{T}( p^{2}) )\, ,
\end{equation}
with  the projection of the Lorentz structure:
\begin{align}\nonumber
  P_{\mu\nu}^{T}(p) = &\,(1-\delta_{\mu 4})(1-\delta_{\nu 4}) \left( \delta_{\mu\nu} - \frac{p_{\mu} p_{\nu}}{\boldsymbol{p}^2} \right)\,, \\[1ex]
  P_{\mu\nu}^{L}(p) = &\,\delta_{\mu\nu} - \frac{p_{\mu} p_{\nu}}{p^2} - P_{\mu\nu}^{T}\,.
\end{align}  
Now the most important upgrade here is a full consideration of  the quark gluon vertex ${\bold \Gamma}^{a}_{\nu}$.   The dressed vertex is in principle  a matrix as:
\begin{equation}
\label{eq:VertexNG}
\begin{gathered}
{\bold \Gamma}^{a}_{\nu} =
\begin{pmatrix}
\Gamma^{a}_{\nu+},& \Xi^{a}_{\nu-}  \\[1.5mm]
\Xi^{a}_{\nu +},& \Gamma^{a}_{\nu-}\\
\end{pmatrix}.
\end{gathered}
\end{equation}

Instead of applying some parametrization to the vertex, here we solve  the Dyson-Schwinger equation for the vertex especially with the non-Abelian diagram  $\mathrm{A}$  in Fig.~\ref{fig:SDE} which is found to be dominant in Ref.~\cite{Gao:2021wun}.   The Abelian contribution $\mathrm{B}$ is $1/N_{c}^{2}$ suppressed in comparison to diagram $\mathrm{A}$ due to  the color matrices.
The DSE for the vertex $\fatg^{a}_{\mu}$ with  the non-Abelian diagram is expressed as
\begin{equation}
	\label{eq:DSEv}
	\fatg^{a}_{\mu}(q,p) =   g_{s} {P}_{\mu\nu}(p-q) \, \fatg^{a,0}_{\nu} + {\bold A}^{a}_{\mu}(q,p)\,,
\end{equation}
with the non-Abelian contributions given by
\begin{align}\label{eq:AandB}
	&{\bold A}^{a}_{\mu}( q,p) =
	{P}_{\mu\nu}(p-q)\sumint_{k} G^{b b^{\prime}}_{\alpha\alpha^{\prime}}((k-q)^{2})G^{c c^{\prime}}_{\beta \beta^{\prime}}((k-p)^{2}) \notag
\\
&\qquad \qquad \quad \times \Gamma^{abc,0}_{\nu\alpha\beta}\fatg^{b^{\prime}}_{\alpha^{\prime}}(k,p) {\bold S}(k) \fatg^{c^{\prime}}_{\beta^{\prime}}( q, k) \,.
\end{align}
In the above formulas, $N_{c} =3$ for SU(3), and
$\Gamma^{abc,0}_{\nu\alpha\beta}$ denotes the classical three-gluon vertex,
\begin{align}
	\Gamma^{abc,0}_{\nu\alpha\beta} =&  f^{abc} g_{s}^{} \bigl[(2k-p-q)_{\nu}^{} g_{\alpha\beta}^{} + (2q-p-k)_{\alpha}^{} g_{\nu\beta}^{} \nonumber\\
& + (2p-q-k)_{\beta}^{} g_{\alpha\nu}^{} \bigr] \,.
\end{align}

The  DSEs in Eqs. \eq{eq:gap1} and \eq{eq:DSEv} are  closed and can be solved numerically. Before doing that, we would like  to give a more direct and simple conclusion by some approximations without losing the reliability.

\emph{Solving the equations in the infrared limit of quark pairing.--}
We   focus on the pseusocalar gap and neglect the momentum dependence  of the mass gap as $\Phi(p)=\Delta\gamma_5 \mathcal{M}$ with $\mathcal{M}$ the color-flavour matrices, and
we simply choose the two flavour superconducting (2SC) channel.   
The choice of the channel  does not change our main conclusion  as it simply changes a common factor   of the self energy which does not have impact on the existence of the pairing gap as will be elaborated below.
The treatment of the momentum dependence here is similar to the NJL model for the dynamical quark mass generation. Such an approximation is effective if one focuses on the non-perturbative properties in the infrared.

Considering the pairing at high $T$ to be small, one may neglect the quadratic contribution of $\Delta$. Such an approximation leads to a simple expression for quark propagator as:
 \begin{equation}
\label{eq:Propa2}
\begin{gathered}
{\bold S}=
\begin{pmatrix}
S_{+}, & S_{+} \gamma_{5} \mathcal{M}S_{-} \Delta^{\ast}  \\[1.5mm]
S_{-} \gamma_{5} \mathcal{M} S_{+} \Delta ,& S_{-}  \\
\end{pmatrix},
\end{gathered}
\end{equation}
moreover, for the chiral symmetric phase above the chiral phase transition, i.e. $T > T_{c}$, one can set the diagonal component in chiral limit as $S_{+} = \gamma_{2} S^{T}_{-} \gamma_{2} = \imag \gamma \cdot p_+ $. If the quadratic contribution is not neglected, the propagator at zero chemical potential will have an additional factor $\frac{p^2}{p^2+\Delta^2}$, which will be applied in the numerical computations.

Now we  discuss about the  dominant structures of the quark-gluon vertex ${\bold \Gamma}^a_\nu$ at $T > T_c$ in the N-G configuration. First, the diagonal parts $\Gamma^{a}_{\nu\pm}$ are constrained from its ``no-pairing'' counterpart. In the quark self-energy, the leading structures are the Dirac term $\gamma_\mu$ and the Pauli term $\sigma_{\mu\nu}(p_\nu-q_\mu)$  as suggested in Refs.\cite{Chang:2013pq,Qin:2013mta,Tang:2019zbk,Lu:2023mkn}.  
%
%
Since the Pauli term is related to the mass function and thus vanishing in the chiral symmetric phase. 
The Dirac term is  constrained via the Slavnov-Taylor identities (STIs)~\cite{Gao:2017tkg,Gao:2021wun}.  This implies that the diagonal part above $T_{c}$ can be  considered as:
\begin{equation}\label{eq:diagvtxTch}
  \bold\Gamma^{a}_{\nu}(p,q) = \frac{ \bold \Lambda^a}{2} F((p-q)^{2}) \gamma_{\nu},
\end{equation}
with $F((p-q)^{2})$ the ghost dressing function.

For the off-diagonal part $\Xi^a_{\nu\pm}$, the Dirac term $\gamma_\mu$ must be vanishing with its coefficient $t_1=0$, which directly from  the constraints of the STIs for quark gluon vertex.
For chiral symmetry breaking structures in $\Xi^a_{\nu}$, one   takes the  Pauli term which is supposed to be  the dominant structure in analogy to the diagonal part. 
The color-flavour structure of $\Xi^a_{\nu}$  is constrained by the STI as:
\begin{align}\label{eq:offdiagcf}
  \mathcal{K}_{+}^{a} &= \frac{1}{2}[ (\lambda^a)^T \mathcal{M} - \mathcal{M} \lambda^a ], \\[1mm]
  \mathcal{K}_{-}^{a} &= \frac{1}{2}[ \lambda^a \mathcal{M} - \mathcal{M} (\lambda^a)^T ].
\end{align}
Note that the anti-symmetric form is taken as that corresponds to the transversal contribution of the vertex DSE in Eq. \eq{eq:AandB}.
This then completes the discussion of the vertex structures and we have:
\begin{equation}\label{eq:Paulivtx}
  \Xi^{a}_{\mu\pm}(p,q) = t_4(p,q) \, \mathcal{K}^{a}_{\pm} \sigma_{\mu\nu} (p-q)^{\nu}\gamma_5, \quad a=1,2,3.
\end{equation}
The definition of $\mathcal{M}$ and the respective color-flavour structure of $\mathcal{M}$ and $\mathcal{K}^{a}_{\pm} $ will be applied in the DSEs in the following and the details  are put in the supplemental material. Similar to the gap, here we  apply the infrared limit approximation for $t_4$ as well with vanishing spatial momentum and  the zeroth Matsubara frequency only:
\begin{equation}\label{eq:momapprox}
t_4(p,q)=t_4(\vec{\,q} , \vec{\,p} = 0;
  q^{+}_{4}, p^{+}_{4} = \pi T + \imag \mu )=t_4\,.
\end{equation}

\begin{figure}[t]
\includegraphics[scale=0.13]{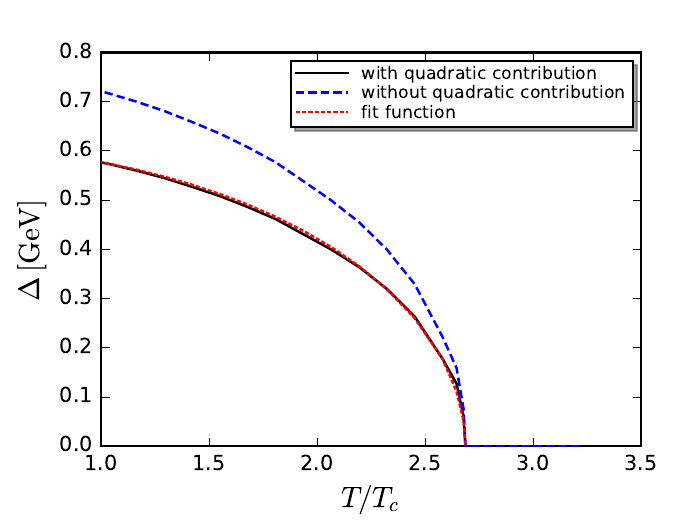}
\vspace*{-6mm}
\caption{The calculated temperature dependence of the quark pairing above the chiral phase transition with (\emph{solid}) and without (\emph{dashed}) the quadratic contribution of $\Delta$. At $T_{\Delta} = 416\,\textrm{MeV}$, the quark pairing vanishes and the temperature dependence near below $T_{\Delta} $ satisfies the fit function as  $\Delta^{2}/0.37[\mathrm{GeV}^{2}] = 1-(T/T_{\Delta})^{2.16}$. }
\label{fig:Delta}
\end{figure}

With the above setup, the DSE of the quark-gluon vertex in Eq.~\eq{eq:DSEv} can be greatly simplified. After tracing out all the color and Dirac matrices and applying \Eq{eq:momapprox}, the DSE yields for the off-diagonal part $ \Xi^{a}_{\mu\pm}$:
\begin{align}
t_{4} & = Z_{1} \Delta + Zt_{4}^{\,2} \Delta , \label{eq:offvtxDSE} \\[1mm]
Z_1 & =\frac{4g^{2}}{9}\sumint_{k} \frac{\vec{k}^{2}}{\bar k_{}^{2} }F^2(\bar k^2)G_{L}(\bar k^2)(G_{L}(\bar k^2)+2 G_{T}(\bar k^2)),\notag \\
Z & =\frac{8g^{2}}{3}\sumint_{k} \frac{\vec{k}^{2} ({k}^{2} + \mu^{2} )}{k_{+}^{2} k_{-}^{2}} G_{T}(\bar{k}^{2})(2G_{L}(\bar{k}^{2}) + G_{T}(\bar{k}^{2}) ) . \quad \notag
\end{align}

\begin{figure*}[t]
\includegraphics[scale=0.13]{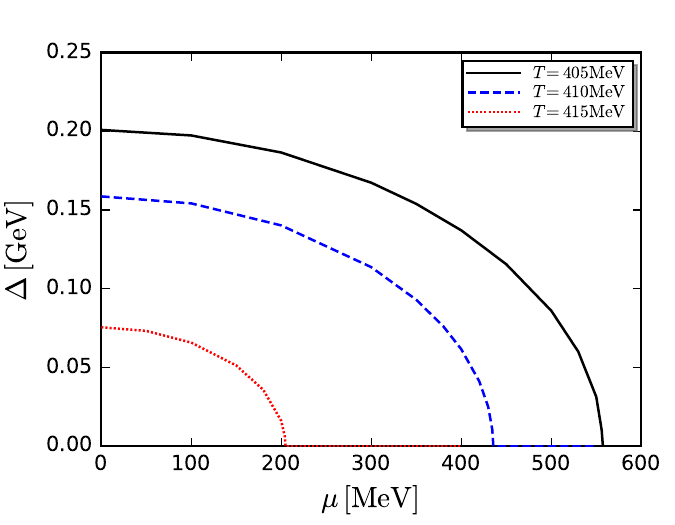}
\includegraphics[scale=0.13]{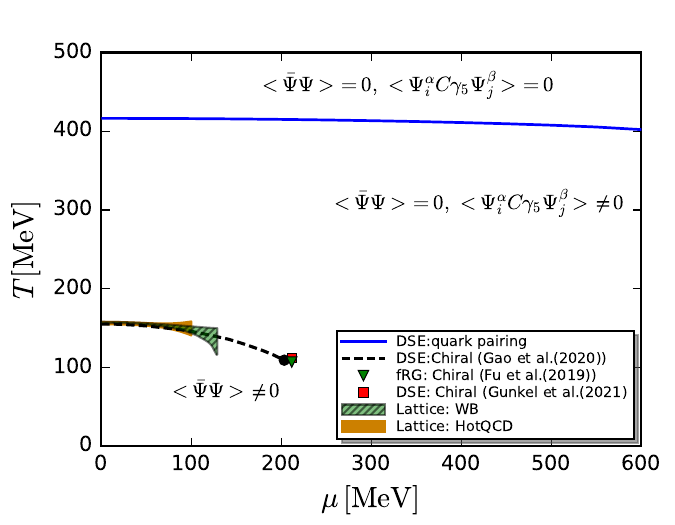}
\caption{The obtained quark chemical potential $\mu$ dependence of $\Delta$ at several temperatures (\emph{left panel}), and the respective phase transition line as the solid line (\emph{right panel}).  For comparison,
the dashed line in the phase diagram is the chiral phase transition line from Ref.~\cite{Gao:2020qsj} together with the critical end point from other functional QCD calculations~\cite{Fu:2019hdw, Fischer:2018sdj}
 and the bands are the lattice QCD results from Ref. \cite{Borsanyi:2020fev} (WB) and Ref. \cite{Bazavov:2018mes} (HotQCD).  }
\label{fig:pt}
\end{figure*}

Note in ultraviolet region with  for instance $p\rightarrow 0$,  the term   $Z_{1} \Delta$  is dominant as $Z_{1}$ is  proportional to  $1/p^{2}$ which then leads to $t_{4} \sim\Delta/p^{2}$. 
Here  taking  the infrared limit as  above, $Z_{1}$ and $Z$ are finite constants. Considering $\Delta$ to be small, the two solutions become 
$t_4=Z_1 \Delta$ and $ t_4=\frac{1}{Z \Delta}$.
The first solution is trivial that leads to the conventional color superconductor which will be discussed later and here we focus on the second solution  which gives:
\begin{equation}
\label{eq:dgvertex}
  \Xi^{a}_{\mu\pm}(p,q) = \frac{1}{Z\Delta}\, \mathcal{K}^{a}_{\pm} \sigma_{\mu\nu} (p-q)^{\nu}\gamma_5, \quad a=1,2,3.
\end{equation}
After putting Eq.~\eq{eq:dgvertex} in the gap equation in Eq.~\eq{eq:gap1}, one has directly for the off-diagonal part $\phi_+$:
\begin{eqnarray}
\label{eq:pairgap}
&&\Delta = -\delta_{m} \Delta + \frac{K}{Z\Delta},\\
&& K = \frac{3}{2} g^{2} \sumint_{k} \frac{k^{+}_{4}\bar k_{4} + \vec{\,k}^{2}}{k_{+}^{2}}(G_{L}(\bar{k}^{2}) + 2G_{T}(\bar{k}^{2})), \label{eq:ApairCond}\notag\\
&& \delta_{m} = \frac{2}{3} g^{2} \sumint_{k} \frac{k^{2}_{4} + \mu^{2} - \vec{k}^{2}}{k_{+}^{2} k_{-}^{2}} F(\bar{k}^{2})(G_{L}(\bar{k}^{2}) + 2G_{T}(\bar{k}^{2})) . \notag
\end{eqnarray}

This gives  a simple expression for the  pairing gap as: 
\begin{equation}   \label{eq:gapsolution} 
\Delta = \sqrt{\frac{K}{Z(1+\delta_m)}} \, .
\end{equation}
Therefore, if $\frac{K}{Z(1+\delta_m)}>0$, one can obtain a  finite solution for $\Delta$ from Eq.~\eq{eq:gapsolution}, 
else if  $\frac{K}{Z(1+\delta_m)}<0$, Eq.~\eq{eq:offvtxDSE}  and Eq.~\eq{eq:pairgap} only allow the trivial solution as $\lambda_{4} =\Delta=0$.

One may further expand $K$ in Eq.(\ref{eq:ApairCond}) as:
\begin{align}
  K &= \frac{3}{2}\langle g^2 A^2 \rangle - \frac{3}{2} \langle g^2 \frac{k^+_4 {p}^+_4}{k_+^2} (G_L(\bar{k}^2)+2G_T(\bar{k}^2)) \rangle,
\end{align}
where the first term on r.h.s is the dimension-2 gluon condensate without the color matrices. The condensate  contains a quadratic divergence that  is artificially induced due to   the neglect  of the momentum dependence of $t_{4}$. 
A complete computation with considering the momentum dependence of $\Delta$ and $t_4$ is in progress, and our preliminary results show that after incorporating the momentum dependence as $\Delta(p^2)$ and $t_4(p^2,q^2)$,
a finite quark pairing gap can still be generated without  the bothering of the divergence. The further  investigations in the whole $T$-$\mu_B$ plane  will be done in our future work.

In conclusion, we show that a  quark pairing  is generated model independently  in the strong QGP phase
In the following, we will compute it numerically   and show the phase diagram with this new paring phase.

\emph{Numerical results for the new phase structure.--}
Here we implement the minimal scheme of DSEs~\cite{Lu:2023mkn} for the gluon propagator which combine the results of the 2+1 flavour gluon propagator at finite temperature from lattice QCD simulation and the quark loop correction in the self energy. 
Besides, to avoid dealing with the quadratic divergence in gluon condensate, we take the vacuum result from lattice QCD  and simply compute its difference from the finite $T$ and $\mu$  as:
\begin{align}
&\langle g^{2} A^{2} \rangle (T,\mu=0) = \langle g^{2} A^{2} \rangle_{vac} + \delta_{ \langle A^{2} \rangle },\\
&\delta_{ \langle A^{2} \rangle }= \int \frac{d^{3} \vec{k}}{(2\pi)^{3}}\left\{T\sum_{n} g^{2} G_{L}(\bar k^{2}) - \int \frac{d k_{0}}{2\pi} g^{2} G_{vac}( \bar k^{2}) \right\}.\notag
\end{align}
We apply $\langle g^{2}A^{2} \rangle_{vac} = 4.4/(N_{c}^{2} -1)\, \textrm{GeV}^{2}$~\cite{RuizArriola:2004en}, and then get a finite quark pairing at temperature above $T=T_{c}$ and vanishes at $T_{\Delta} = 416\,\textrm{MeV}$. 
The obtained temperature dependence of quark pairing  is depicted in  Fig.~\ref{fig:Delta}. The quark pairing gap   supports a  second order phase transition at temperature $T_\Delta$, as one has $\Delta=0$ above $T_\Delta$, and   near below $T_{\Delta}$:
\begin{equation}
 \Delta^2 \propto 1- ( T/T_{\Delta})^{a},
\end{equation}
with the best fit of the power law  as $a=2.16$. If considering $\Delta$ as the order parameter, the relation then yields a mean filed critical exponent as $\beta=1/2$. 

One can include the quadratic contribution of $\Delta$ by adding the factor $\frac{k^2}{k^2+\Delta^2}$ in the integrand of $Z_1$, $Z$, $\delta_m$ and $K$ and numerically solve Eq. \eq{eq:offvtxDSE} and Eq. \eq{eq:pairgap} together with the above treatment for gluon condensate. As depicted in Fig.~\ref{fig:Delta}, the quadratic contribution makes the gap decrease slightly but it does not affect $T_\Delta$ as the gap is vanishing there.

For finite chemical potential, the condensate can be computed through the $\mu$-derivative as:
\begin{align}
&\langle g^{2}A^{2}  \rangle (T,\mu) = \langle g^{2}A^{2} \rangle (T,\mu=0) + \int_{0}^{\mu} d\mu \chi_{g} \,,\\
&\chi_{g} = \frac{\partial  \langle g^{2} A^{2} \rangle (T,\mu)}{\partial \mu} = -\sumint_{k} g^{2} G_{L}^{2} (\bar k^{2})\frac{\partial m^{2}_{g}}{\partial \mu},\notag
\end{align}
where the thermal mass  $m^{2}_{g}$ can be taken as the hart thermal loop value for the temperature domain we considered here,  
and one has  $\partial m^{2}_{g}/\partial \mu = N_{f}\mu g^{2}/\pi^{2}$ with $N_{f} =3$.

At finite $\mu$, the gap $\Delta$ is in general complex similar to the mass function. Here we simply compute the average of $\Delta$ with both the first positive and negative Matsubara frequency of $p\,,q$ as ${p}^{\pm}_{4} = {q}^{\pm}_{4} = \pm\pi T + \textrm{i} \mu$. This allows one to define the phase transition taking place at where the average of $\Delta$ vanishes.    
We depict  the $\mu$ dependence of the pairing at several temperatures and finally the respective phase diagram together with the chiral phase transition from the previous studies in Fig.~\ref{fig:pt}. 
In general, the quark pairing exists above $T_{c}$ up to $T\approx 2$ -- $3T_{c}$, which coincides in fact with the previous conjectures of a possible new phase existing at this area~\cite{Glozman:2022lda,Qin:2010pc,Gao:2014rqa,Fischer:2017kbq, Shuryak:2003ty,Shuryak:2014zxa,McLerran:2008ua,McLerran:2007qj}.  The discontinuous derivative of the quark pairing at $T_{\Delta}$ further characterizes  this new phase  that is distinct from the wQGP with quasi quarks and gluons.

\emph{Discussions and outlooks.--}
We firstly  emphasize  that the quark pairing gap is induced purely by the non-Abelian feature of the vertex as it generates the $\frac{1}{\Delta}$--type interaction which induces naturally a finite $\Delta$ in the gap equation in Eq.~\eq{eq:pairgap}. 
This specific form makes the gap robust against the choices of the coupling strength and the channels of the pairing. 
 The pairing is closely related to the gluon condensate and  therefore, its generation mechanism is   dominant 
 by  the glue dynamics. 

One can also check the contribution of the Abelian diagram B in the quark gluon vertex. 
Following similar procedures,  it gives the coefficients that is proportional to $\Delta$ as $t_{4} \propto \frac{g^{2}}{2N_{c}}\Delta$  instead of $1/\Delta$ which only allows the trivial solution, i.e.  the conventional color superconducting phase (CSC). 
In particular,  the gap in  conventional CSC is generated through  the quadratic correction of the pairing  in the denominator of the quark propagator in the form of $\frac{1}{p^2+\Delta^2}$. This type of propagator gives a gap  that is proportional to chemical potential $\mu$  as $\Delta\sim\mu\, e^{-\frac{\mathrm{const}}{g}}$ in weak coupling limit, which its mechanism is distinct from that for the quark pairing pattern found here.

The quark pairing phase has greatly enriched the QCD phase diagram. The pairing  represents  a color deconfined  phase just above the chiral phase transition, 
however,  the quark is  confined into colored bound states which indicates a partial deconfined phase. 
Only after the phase transition at $T_{\Delta} \approx 2$ -- $3T_{c}$, the QCD matter becomes quasi quarks and gluons.  
The quark pairing phase has a large overlap with the chiral spin symmetric phase and the region with zero mode in quark spectrum.  
The bosonic quark pairing and the proposed fermionic field with the string-like interaction in chiral spin symmetric phase are two  collective modes, which are reminiscent of the duality between the pairing and vortex dynamics where the fermion mode emerges from an inhomogeneous pairing as $\Delta(\vec{r})$ ~\cite{Chamon:2010ks,Eto:2013hoa,Chatterjee:2016ykq}. 
A further investigation on their relation and generating mechanism may reveal some new features of QCD matter.    
Moreover, this new phase explains the small shear viscosity and thermodynamic quantities of sQGP. Its exceptional properties will help further to understand the heavy ion collision experimental measurements.


\begin{section}{Acknowledgements}

FG and YL thank Wei-jie Fu and Jan M. Pawlowski 
for the fruitful discussions.
FG is  supported by the National  Science Foundation of China under Grants  No. 12305134. YL and YXL are supported by the National Natural Science Foundation of China under Grants  No. 12247107 and  No. 12175007.
\end{section}

\bibliography{refs}

\renewcommand{\thesubsection}{{S.\arabic{subsection}}}
\setcounter{section}{0}
\titleformat*{\section}{\centering \Large \bfseries}

\onecolumngrid

\section*{Supplemental Material}

\subsection{The color-flavour structures of the quark pairing and the quark gluon vertex}\label{sec:app}

To calculate the color-flavour structures of quark pairing, one needs to combine the color space with the flavour space and apply the basis with $\{(r,u),(g,d),(b,s),(r,d),(g,u),(r,s),(b,u),(g,s),(b,d)\}$ \cite{Buballa:2003qv,Nickel:2006vf,Alford:2007xm,Muller:2016fdr}. Here we consider the two flavour quark pairing, and the pairing is then defined with the matrix representation $\mathcal{M}$ as:

\begin{equation}
\mathcal{M}=\begin{bmatrix}
0&-1&0&0&0&0&0&0\\
-1&0&0&0&0&0&0&0\\
0&0&0&0&0&0&0&0\\
0&0&0&0&1&0&0&0\\
0&0&0&1&0&0&0&0\\
0&0&0&0&0&0&0&0\\
0&0&0&0&0&0&0&0\\
0&0&0&0&0&0&0&0
\end{bmatrix}.
\end{equation}

The color factor for the diagonal part of the DSEs of propagator and quark gluon vertex is straightforward as given in Refs. \cite{Nickel:2006vf,Muller:2016fdr}. For the off diagonal part of $\Xi_\mu^a$, we consider the Slavnov-Taylor identity:
\begin{equation}
   \imag k_{\mu} {\bold \Gamma}^{a}_{\mu} \propto \frac{ \bold \Lambda^a}{2} \, {\bold S}^{-1}(q) - {\bold S}^{-1}(p) \, \frac{ \bold \Lambda^a}{2},
\end{equation}
with ${\bold \Lambda}^a$ defined in \Cref{eq:barevtx} and apply the following color-flavour structure:
  \begin{align}\label{app-eq:offdiagcf}
  \mathcal{K}_{+}^{a} &= \frac{1}{2}[ (\lambda^a)^T \mathcal{M} - \mathcal{M} \lambda^a ], \\
  \mathcal{K}_{-}^{a} &= \frac{1}{2}[ \lambda^a \mathcal{M} - \mathcal{M} (\lambda^a)^T ].
\end{align}

The consistency relation of the color-flavour structure for the quark pair in Eq. \eq{eq:gap2} is then given as:
\begin{equation}\label{app-eq:qpaircf}
  \frac{(\lambda^a)^T}{2} \mathcal{K}_{+}^{a} = \frac{\lambda^a}{2} \mathcal{K}_{-}^{a} = \frac{3}{2} \mathcal{M}.
\end{equation}

Moreover, the loop diagram of the quark gluon vertex  gives the color-flavour structure as:
\begin{equation}\label{app-eq:offdiagcf-self}
 f_{abc} \mathcal{K}_{\pm}^{b} \mathcal{M} \mathcal{K}_{\pm}^{c}=C_{\pm} \,\mathcal{K}_{\pm}^{a}.
\end{equation}
Here  we  only consider the coefficients for $a=1,2,3$ to be non-vanishing so that $C_{\pm} = 2$. This then completes the color factors  involved in this work.

\end{document}